\documentclass[12pt]{article}

\usepackage{epsfig}
\usepackage{url}
\usepackage{hyperref}

\newcommand{\GeV}       {\mbox{GeV}}
\newcommand{\TeV}       {\mbox{TeV}}
\newcommand*{\pt}       {\mbox{$p_\mathrm{T}$}}
\newcommand{\ttbar}     {\mbox{$t\bar{t}$}}

\newcommand{\zprim}     {\mbox{$Z'$}}
\newcommand*{\ttH}      {\mbox{$t\bar{t} H$}}

\def\Title#1{\begin{center} {\Large {\bf #1} } \end{center}}

\usepackage[switch*]{lineno} 

\begin{document}

\Title{Flavour tagging with graph neural networks with the ATLAS detector}

\begin{center}
{\it Arnaud Duperrin\index{Duperrin, A.}, on behalf of the ATLAS Collaboration\footnote{Copyright CERN for the benefit of the ATLAS Collaboration. CC-BY-4.0 license}\\
CPPM, Aix-Marseille Universit\'e, CNRS/IN2P3, Marseille, France. \bigskip

{\small Presented at DIS2023: XXX International Workshop on Deep-Inelastic Scattering and Related Subjects, Michigan State University, USA, 27-31 March 2023.} 
}
\end{center}

Abstract: The identification of jets containing a $b$-hadron, referred to as $b$-tagging, plays an important role for various physics measurements and searches carried out by the ATLAS experiment at the CERN Large Hadron Collider (LHC). The most recent $b$-tagging algorithm developments based on graph neural network architectures are presented. Preliminary performance on Run 3 data in $pp$ collisions at  $\sqrt s = 13.6$~\TeV\ is shown and expected performance at the High-Luminosity LHC (HL-LHC) discussed.

\section{Introduction}
Bottom jets ($b$-jets) originate from the decay of $b$-hadrons. Processes with heavy-flavours quarks ($b$,$c$) play a key role in the LHC physics program like for instance in the \ttH~\cite{ATLAS:2021qou} production mode where the Higgs boson decay into a $b$-quark pair ($H \rightarrow b\bar{b} $) as measured by the ATLAS experiment~\cite{ATLAS:2008xda}. Flavour tagging aims to identify the flavour of a jet ($b$-, $c$-, or light-jet).

The characteristically long lifetime of hadrons containing $b$-quarks of the order of 1.5~ps can be used to identify $b$-jets. A class of algorithms explicitly reconstruct the production position (vertex) of the tracks originating from the $b$-hadron decays which is displaced from the primary interaction point, or exploits the displacement of reconstructed charged particles trajectories (tracks) by measuring their impact parameter. Another class of algorithms, based on neural network architectures, use as inputs the impact parameters and kinematics of the tracks. A recurrent neural network treats track collections as a sequence while a deep sets model has a permutation-invariant and highly parallelisable architecture.

Current ATLAS flavour tagging algorithms rely on the outputs on these taggers that are then combined using
machine learning techniques forming the so-called DL1 algorithm series. The following references provide more details about the DL1r~\cite{ATLAS:2022qxm} and DL1d~\cite{ATL-PHYS-PUB-2022-047} taggers. Considerable improvements in performance are obtained over previous generations of taggers which were based on boosted decision trees or likelihood discriminants.

Recently, ATLAS released an improved tagger based on graph neural networks, name GN1~\cite{ATL-PHYS-PUB-2022-027}. 

\section{Graph neural network jet flavour tagging}

A graph represents the relations (edges) between a collection of entities (nodes). The GN1 tagger is a new approach which utilizes a graph neural network to predict the jet flavour directly taking as inputs the individual tracks parameters and their uncertainties together with the jet \pt\ and $\eta$. Each node in the graph corresponds to a single track in the jet, and is characterised by a feature vector (or representation) of length 23 based on above inputs. A fully connected graph network architecture between nodes is used.

The graph is trained with two auxiliary objectives to aid the primary objective of the jet flavour identification. The first one performs track-pair vertex compatibility (i.e.if the two tracks in the pair originated from the same point in space) removing the need for inputs from a dedicated secondary vertexing algorithm. The second auxiliary objective predicts for each track within the jet the underlying physics process from which each track originated (i.e. whether it's a $b$, $c$, light, pile-up, fake track etc.). 

The training for the primary and auxiliary objectives uses truth information available only in simulation in addition to reconstructed quantities (i.e. tracks, jets) available in both collision data and simulation. To train and evaluate the model, simulated Standard Model \ttbar\ and beyond Standard Model \zprim\ resonances decaying to heavy flavour quarks events are used. 

\section{Performance of the GN1 tagger}

\begin{figure}
\begin{center}
\centering
\includegraphics[width=0.9\linewidth]{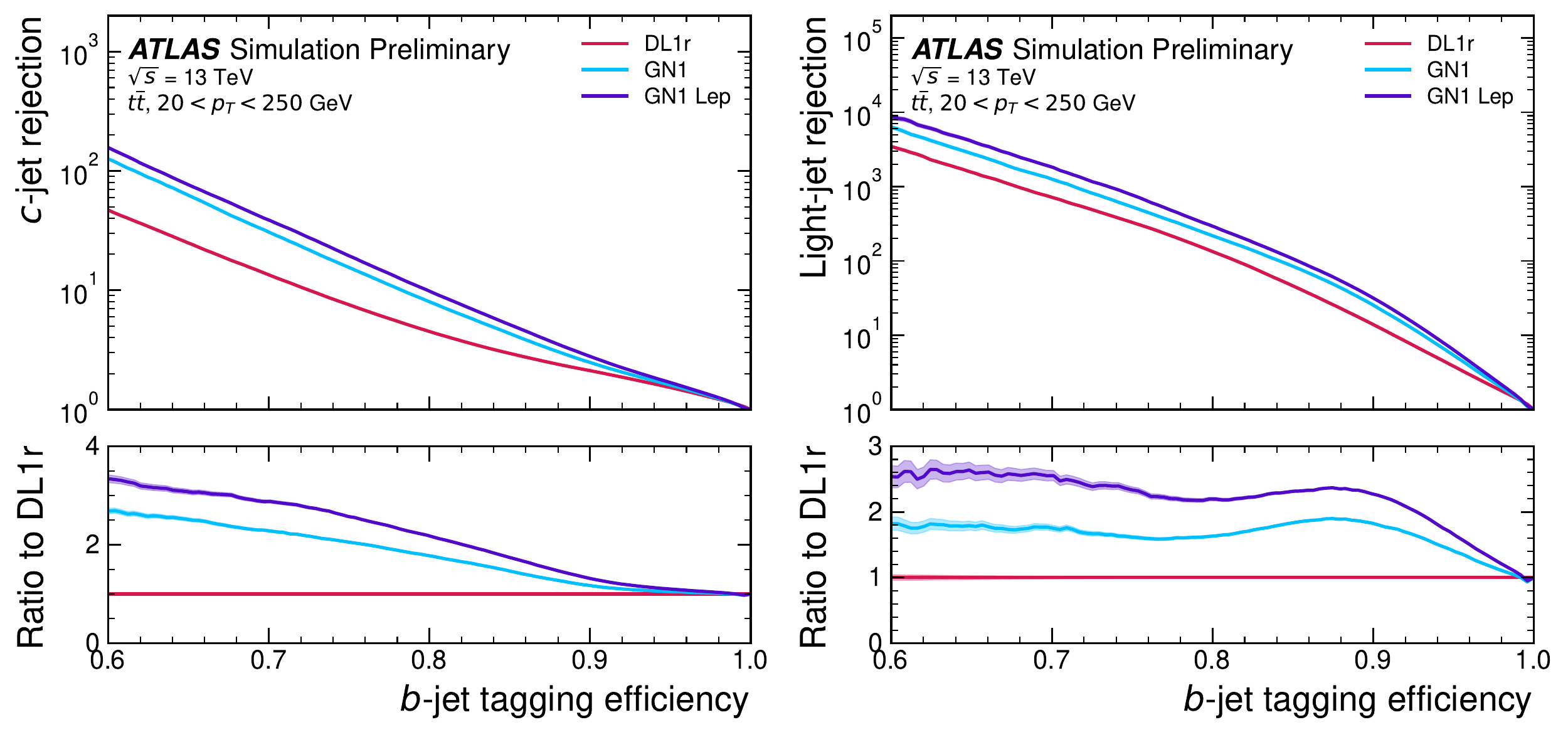}
\caption{$c$-jet (left) and light-flavour jet (right) rejections of the GN1~\cite{ATL-PHYS-PUB-2022-027} graph neural tagger as a function of the $b$-jet tagging efficiency for jets in a \ttbar\ simulated sample with $\pt > 20$ \GeV. The ratio with respect to the performance of the DL1r algorithm~\cite{ATLAS:2022qxm}, used for previous flavour-tagging studies, is shown in the bottom panels. The GN1 Lep variant includes an additional track-level input which indicates if the track was used in the reconstruction of an electron or a muon.}
\label{fig:perf-gn1-ttbar}
\end{center}
\end{figure}

The performance of GN1 is shown in Figure~\ref{fig:perf-gn1-ttbar} in a \ttbar\ sample demonstrating considerably better $c$- and light-jet rejection compared with the DL1r tagger across the full range of $b$-jet tagging efficiencies probed. For instance, at $70\%$ $b$-jet tagging efficiency, the $c$-jet rejection improves by a factor of $\sim2.1$ and the light-jet rejection improves by a factor of $\sim1.8$ with respect to DL1r. For high-\pt\ jets in a \zprim\ sample with $250 ~\GeV < \pt < 5000$ \GeV, at $30\%$ $b$-jet tagging efficiency, the $c$-jet (light-jet) rejection improves by a factor of $\sim2.8$ ($\sim6$).

Auxiliary objectives help the jet flavour prediction via a supervised attention mechanism. An attention mechanism is a way of learning which parts of the data are more important than others. In the context of the GN1 tagger, the model learns to pay more attention to tracks from heavy flavour decays. In addition, it helps with the interpretability of the network by providing more detailed information about how tracks are classified and about their vertex compatibility for a each jet. GN1 correctly identifies $80\%$ of truth vertices inside $b$-jets for instance.

\begin{figure}
\begin{center}
        \includegraphics[width=0.47\linewidth]{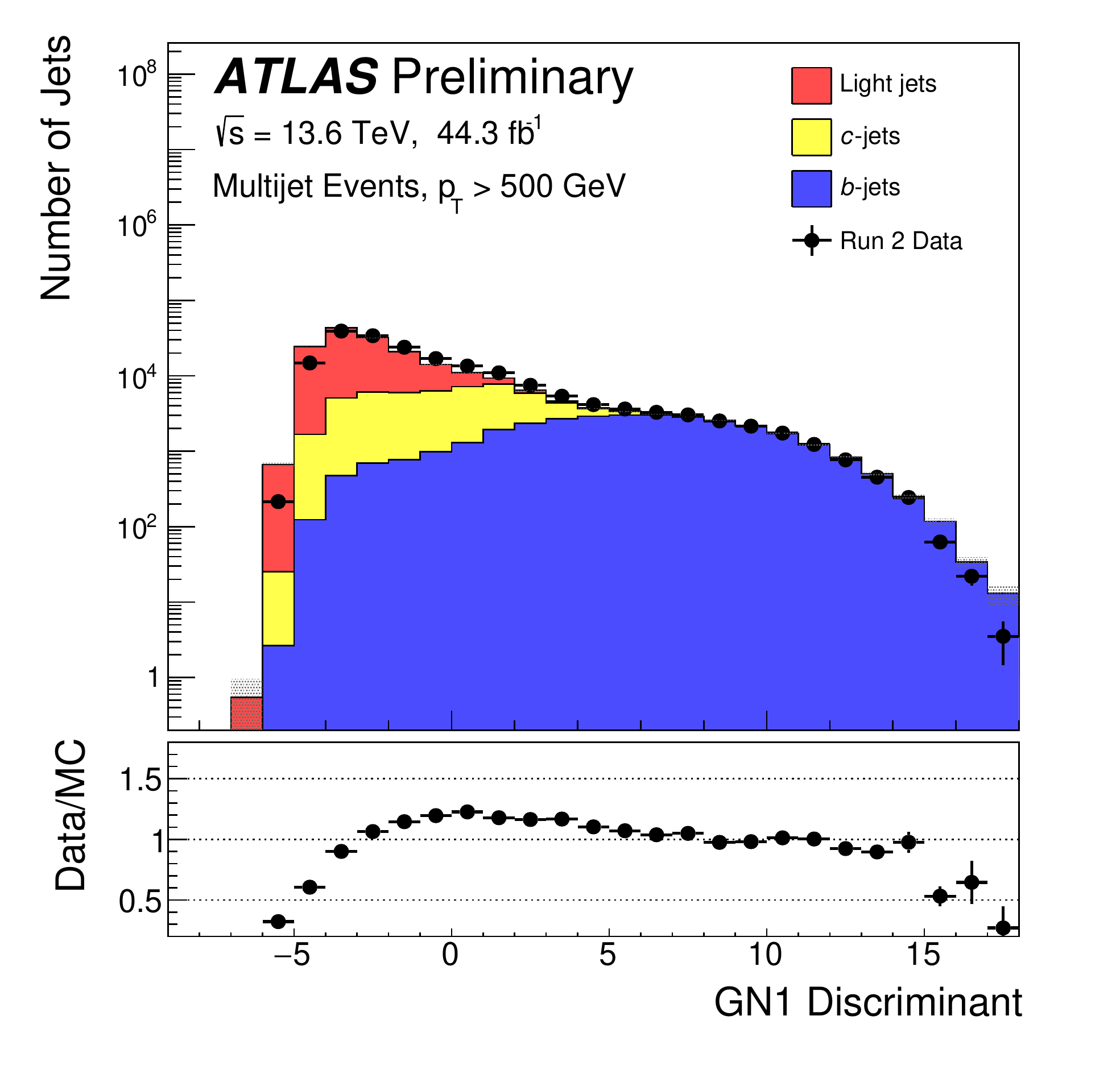}
        \includegraphics[width=0.47\linewidth]{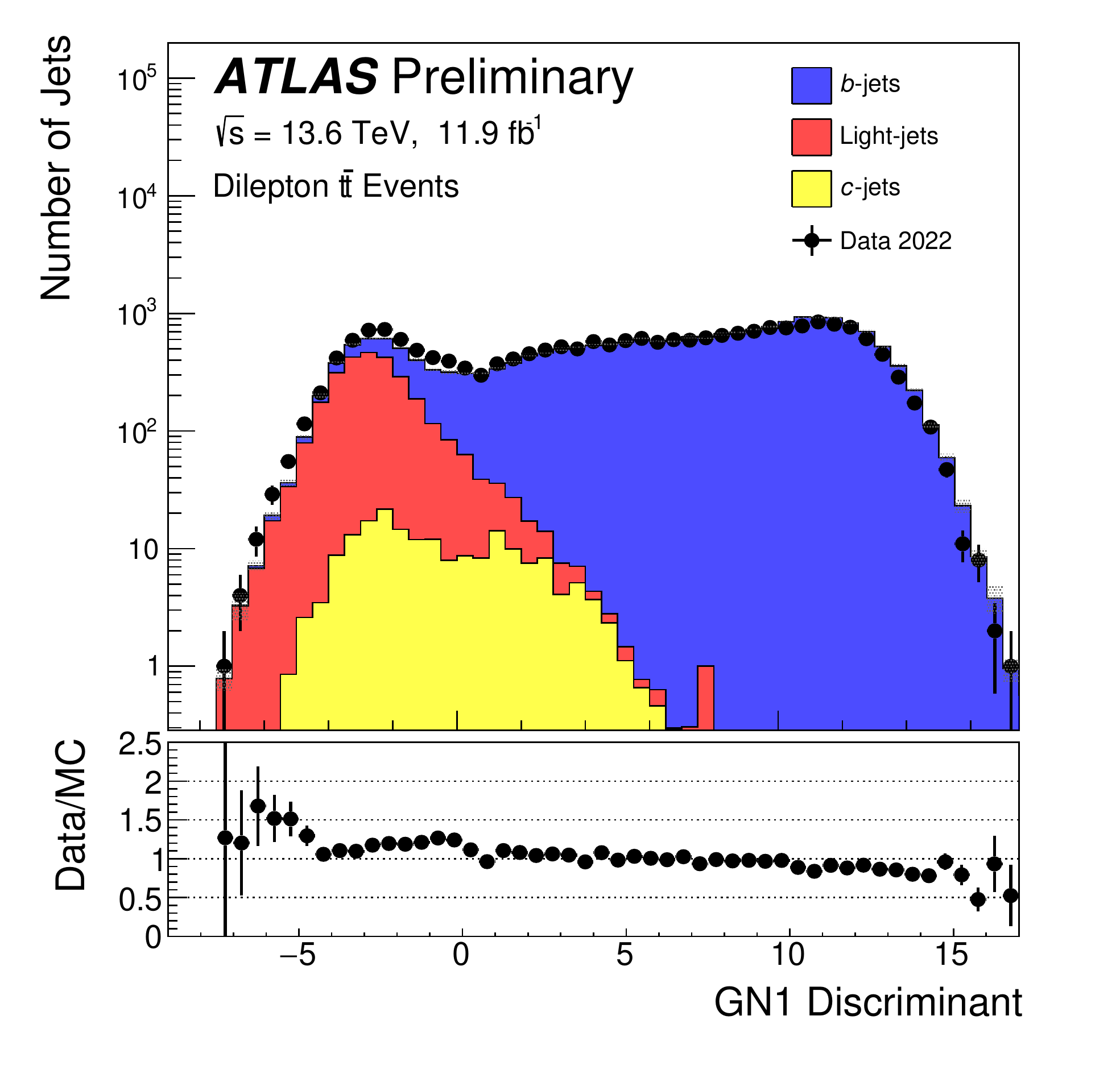}
    \caption{Tagging algorithm output distributions for GN1 in multijet events (left) and \ttbar\ dileptons events (right). The simulation is scaled to match the total yield in Run 3 data. The shaded bands refer to the statistical uncertainties on the simulation. More details about the selections in Ref.~\cite{GN2publicPlots}.}
    \label{fig:gn1-discr-run3data}
\end{center}
\end{figure}

The agreement of the GN1 discriminant with Run 3 data is displayed in Figure~\ref{fig:gn1-discr-run3data} in multijet and \ttbar\ dileptons events. A good agreement is observed from these preliminary comparisons, in particular in the region where $b$-tagging operating points are defined for analyses (positive value of the discriminant).   

The GN1 model is tested on other MC samples to check if it is learning generator-dependent information. The overall dependence is found to be of the order of O(3\%) for $b$-jets and O(6\%) for $c$-jets corresponding to similar values obtained with previous generation DL1r/DL1d taggers.

\section{GN1 at HL-LHC}

The upcoming High-Luminosity LHC upgrades are expected to be completed by 2029 to operate at an average number of collisions per bunch crossing of up to 200 compared to 55 during Run 3 making $b$-tagging even more challenging. A significant upgrade of the tracking detector with a new all-silicon Inner Tracker (ITk~\cite{ATL-PHYS-PUB-2022-047}) will be greatly beneficial to flavour tagging by guaranteeing tracking performance at least equivalent to what is currently achieved with the Run 3 detector and extending the coverage up to $|\eta|=4$. The GN1 improvements~\cite{ATL-PHYS-PUB-2022-047} evaluated with respect to previous generations of flavour tagging algorithms (also tuned to HL-LHC conditions) are, for instance, up to 30\% in $b$-efficiency at high-\pt\ and 15\% in the forward region ($|\eta|>2.5$).

\section{Pushing further improvements (GN2)}

Building upon the success of GN1, recent developments have extended its features leading to the GN2 tagger where the majority of the changes are optimisations for the model hyperparameters. The difference between GN1 and GN2 is sumarized in Table~\ref{tab:gn2-hyperpara}. The learning rate is based on the One-Cycle learning rate scheduler~\cite{2018arXiv180309820S} and the model follows the transformer architecture~\cite{2017arXiv170603762V}. The attention type has been changed with no effect on physics performance but it improves the training time and memory footprint. GN2 separates the computation of the attention weights from the computation of the updated node representations and uses a dense layer in between the attention layers. The training statistics were increased from 30 million jets to 192 million training jets. For a $b$-jet efficiency of 70\%, the light ($c$)-jet rejection is improved by a factor of 2 (1.5) for jets coming from \ttbar\ decays with transverse momentum $20 ~\GeV < \pt < 250 ~\GeV$. For jets coming from \zprim\ decays, the light ($c$)-jet rejection improves by a factor 1.2 (1.75) at 30\% $b$-jet efficiency.

\begin{table}
\begin{center}
\caption{Changes between GN1~\cite{ATL-PHYS-PUB-2022-027} and GN2~\cite{GN2publicPlots}.}
\begin{tabular}{c}
        \includegraphics[width=.86\linewidth]{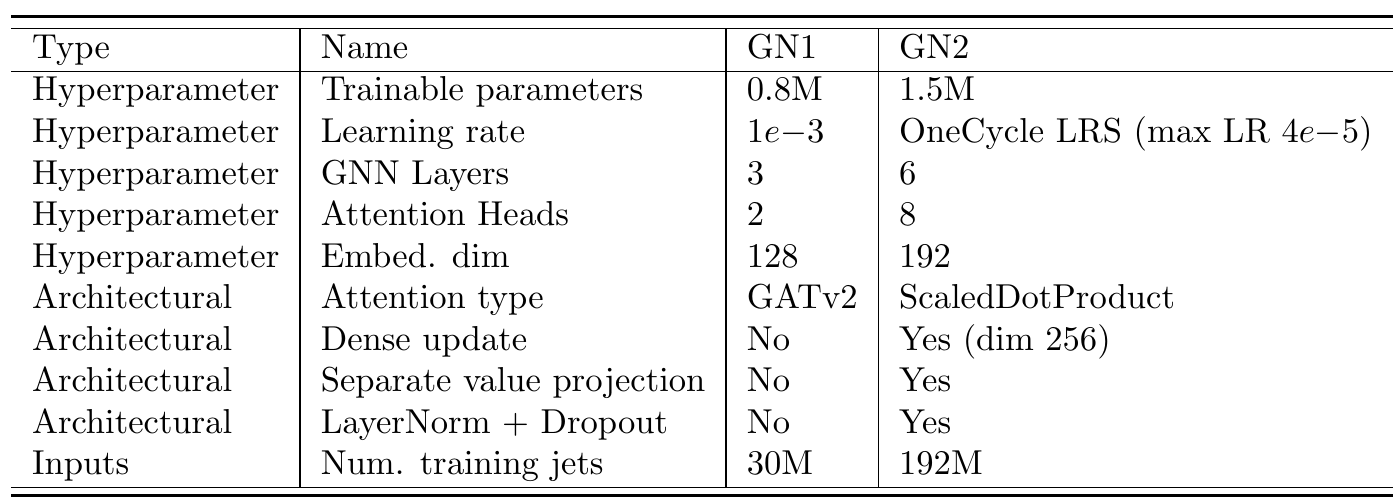}
\end{tabular}

\label{tab:gn2-hyperpara}
\end{center}
\end{table}

\section{Conclusion}

The next generations of ATLAS $b/c$ taggers (GN1/GN2) are based on graph neural networks models. They show very promising results with a factor of four improvement in background rejection with respect to the DL1 tagger series. Checks on collision data have been performed and the Collaboration is now moving towards full calibration. From the results presented, strong benefits on the ATLAS physics program at Run 3 LHC and HL-LHC are expected.



\bibliographystyle{hep}
\bibliography{article}

\end{document}